\documentclass[11pt,preprint2]{aastex}

\shorttitle{A Search for Host Galaxies of 24 GRBs}
\shortauthors{Ovaldsen et al.}

\begin{document}

\title{A Search for Host Galaxies of 24 Gamma Ray Bursts}


\author{J.-E.~Ovaldsen\altaffilmark{1}, A.~O.~Jaunsen\altaffilmark{1},
  J.~P.~U.~Fynbo\altaffilmark{2}, J.~Hjorth\altaffilmark{2},
  C.~C.~Th{\"o}ne\altaffilmark{2}, C.~F{\'e}ron\altaffilmark{2},
  D.~Xu\altaffilmark{2}, J.~H.~Selj\altaffilmark{1},
  J.~Teuber\altaffilmark{1,2}}

\altaffiltext{1}{Institute of Theoretical Astrophysics, University of
  Oslo, P. O. Box 1029, Blindern, N-0315 Oslo, Norway}
  \email{j.e.ovaldsen@astro.uio.no}
\altaffiltext{2}{Dark Cosmology Centre, Niels Bohr Institute,
  University of Copenhagen, Juliane Maries Vej, DK-2100 Copenhagen
  {\O}, Denmark}


\begin{abstract}
  We report the results from observations of 24 gamma ray burst (GRB)
  fields from 2005 and 2006 undertaken at the Danish 1.54m telescope
  at ESO/La Silla.  Photometry and positions for two previously
  unpublished host galaxy candidates (GRBs 050915 and 051021) are
  presented, as well as for eight other detected objects which are
  either known GRB hosts or candidate hosts. The candidates are
  suitable for spectroscopic follow-up in order to have their
  redshifts and other physical characteristics determined. In the
  cases where no likely host candidate is detected inside the refined
  {\it Swift} XRT error circle we are still able to put interesting
  and rather deep limits on the host magnitude. Based on our
  detections and upper limits we have performed simulations which
  suggest that the host galaxies are drawn from a fainter sample than
  previous (i.e.\ pre-{\it Swift}) studies.

\end{abstract}


\section{Introduction}
The properties of galaxies hosting long gamma ray bursts (GRBs) were
early on realized to give important clues on the nature of GRBs
\citep{Fenimore93,Band98,Hogg99}. Furthermore, it was also realized
early on -- in the so-called ``afterglow'' era, when the cosmological
distance scale for long GRBs was established -- that GRBs may be ideal
tracers of (massive) star-formation throughout the observable
Universe. Hence, a well selected sample of GRB host galaxies could be
used to characterize the properties of the various types of galaxies
that dominate the cosmic star-formation as a function of cosmic epoch
\citep{Totani97,Wijers98,Mao98,Blain00}.

Prior to the currently very successful {\it Swift} satellite
\citep{Gehrels04} there were a number of published host studies
\citep[e.g.,][]{Hogg99,LeFloch03,Christensen04,Fruchter06}. The main
conclusion based on these and other pre-{\it Swift} studies were that
the hosts of GRBs are typically faint, star-forming dwarf galaxies.
The median magnitude of pre-{\it Swift} hosts is about $R=25$.
However, these conclusions are based on incomplete samples as
typically only about 30\% of pre-{\it Swift} GRBs are sufficiently
well localized to allow an unambiguous host identification
\citep[e.g.,][]{Fynbo01,Berger02}. Therefore, these conclusions may
not be valid for GRB hosts in general.

With the {\it Swift} satellite the rate of well localized long GRBs
has gone up by about an order of magnitude. Furthermore, given the
very rapid and precise localization, thanks primarily to the X-ray
telescope (XRT) on {\it Swift}, the fraction of {\it Swift} GRBs
without sub-arcsecond localization is now significantly smaller, and
even based on the (refined) XRT positions alone it is possible to make
interesting conclusions on the properties of the hosts.

So far no systematic study of the hosts of {\it Swift} GRBs has been
published. We here make a first attempt to characterize a sample of
hosts related to {\it Swift} GRBs based on relatively deep images
collected at the Danish 1.54m telescope on La Silla in Chile.  The
paper is organized as follows: We describe the observations and data
set in Section~\ref{S:obs_data} and the photometric reductions in
Section~\ref{S:reductions}. The results for our 24 GRB fields (22 {\it
  Swift} and two {\it HETE} bursts) are presented in
Section~\ref{S:results}; we derive upper limits for the brightness of
the host for each GRB and provide positions and magnitudes for the
detected host galaxy candidates.
In Section~\ref{S:comparison} our photometric results for a subsample
of 15 GRB fields -- i.e.\ long {\it Swift} GRBs with confirmed optical
transients (OTs) -- are compared to the pre-{\it Swift} host sample of
\citet{Fruchter06}. Our results give tentative evidence that GRB host
galaxies from the post-{\it Swift} era are fainter than those
belonging to earlier host studies. Section~\ref{S:summary} summarizes
our main findings.


\section{Observations and data set} \label{S:obs_data}
The data were obtained with the Danish 1.54m telescope at the European
Southern Observatory on La Silla, Chile, 2005 October 24 to November
21 and 2006 February 19 to March 15. The CCD detector on the DFOSC
(Danish Faint Object Spectrograph and Camera) instrument has a current
default readout area of 2148 $\times$ 2102 pixels, a pixel scale of
0.395 arcsec pix$^{-1}$, a gain of 0.74 electrons ADU$^{-1}$, and a
readout noise of 3.1 electrons.

We selected our GRB targets (Table~\ref{t:grb_info}) mainly based on
observability at the two epochs of observations. Priority was given to
favorably placed GRB fields, so high ($>2$) airmass targets were in
some cases skipped. A discussion of the completeness of our sample can
be found in Section~\ref{S:comparison}.
Images were taken in the Bessel $B$-, $V$-, and $R$-band, as well as
Gunn $i$-band.  We mostly took 600 s exposures which later were
combined and stacked; the first four columns of Table~\ref{t:obs}
summarizes the data set and quote total exposure time and average
seeing values in the different filters for the 24 GRB fields in this
sample. Unless otherwise noted, all observations were performed well
after the optical afterglow 
had faded.


\section{Reductions} \label{S:reductions}

\subsection{Pre-processing and stacking}
The images were pre-processed in the standard way using IRAF (Image
Reduction and Analysis Facility). The bias level was subtracted using
the overscan region on each science frame and a master bias frame
computed from typically 17--20 individual frames on each night.
Flat-fielding was performed exclusively by means of sky flats. Dark
current did not need to be corrected for as the dark level was found
to be totally negligible. Images taken in the $i$-band (and to some
extent the $R$-band) were affected by fringing. 
We tried to minimize the effect on the photometry by carefully
selecting the region from which the local sky level was estimated. The
$i$-band images also suffered from a large gradient in the background
level across the CCD frame (from upper left to lower right corner),
again making careful local background estimation important.

Individual frames were stacked in IRAF with offsets according to the
World Coordinate System (WCS) keywords in the image headers. However,
the original WCS information was incorrect and had to be re-computed
first. We used SExtractor \citep{sex96} for source detection on the
individual images and then correlated the pixel positions with sky
coordinates from the USNO-B1.0 Catalog\footnote{The accuracy of the
  USNO-B1.0 Catalog is 0$\farcs$2.} \citep{Monet03} using the
\texttt{imwcs} program in
WCSTools.\footnote{\url{http://tdc-www.cfa.harvard.edu/software/wcstools/}}
The WCS fit was typically derived from 100--200 objects in the
generous 13.7$\arcmin$ field of view and allowed for a robust
astrometric calibration. Images with high sky values and/or short
exposure times were down-weighted during the stacking. Specifically,
in IRAF's \texttt{IMCOMBINE} task the frames were scaled and weighted
by the exposure time, and the zero level offset was taken as the mode
within a central region of the CCD frame.

\subsection{Photometry} \label{S:phot}
For photometry we used an IDL (Interactive Data Language) code
\citep[see e.g.][]{Ovaldsen03a, Ovaldsen03b} featuring circular or
elliptical apertures with correction for fractional pixels at the
aperture border. Centering was performed using paraboloidal
representation of the nine central pixel values of the source in
question \citep{Teuber03}. The local background level was calculated
from several apertures arranged around the object of interest, and any
apertures containing cosmics, sources etc.\ were automatically
discarded.

The photometric calibration was performed relative to the Landolt
($BVR_CI_C$) system. Zero-points, extinction coefficients and color
terms were derived from observations of Landolt standard fields on two
nights with photometric conditions. Color terms were negligible for
all filters, except $B$ (the color-correction term, $c_{B}(B-V)$, had
$c_{B}= -0.12$).  Each GRB field was calibrated using three to five
comparison stars.

We quote a detection limit, or upper limit, for all GRBs in the
sample.  This parameter is a function of (at least) background noise
and seeing. 2$\sigma$-limits were calculated in the following way:
\begin{equation}
\mathrm{Mag}(2\sigma) = \mathrm{ZP} - 2.5\log\left(
2 C \sqrt{\pi r_d^2 \sigma_B^2} \right) ,
\end{equation}
where ZP is the photometric zero-point, $C$ is the aperture
correction, $r_d$ is the radius of the circular detection aperture
(see below), and $\sigma_B$ is a conservative measure of the
background noise, i.e.\ the standard deviation of the sky values. $C$
is the ratio between the total flux of the reference stars and the
flux inside the detection aperture whose radius, $r_d$, is set to 1.5
times the half width at half maximum (HWHM). The HWHM was measured
from the point-spread function (PSF) on each image.
The factor two inside the parenthesis corresponds to a 2$\sigma$-limit.

\section{Results} \label{S:results}

The stacked images of the 24 GRB fields were examined with respect to
objects in or near the {\it Swift} XRT error circle and other reported
positions, i.e.\ from afterglow observations in various bands mainly
from ground-based telescopes.
We use the new, refined {\it Swift} XRT positions and errors presented
in \citet{Butler06}, which are determined by matching X-ray field
source positions directly to those of counterpart optical sources.
We also note that several {\it Swift}-UVOT (UV/Optical Telescope)
positions for the afterglows are inconsistent with the refined XRT
error circle and OT positions from other groups.  A detection limit,
i.e.\ upper limit for the brightness of the host, was derived for each
GRB field, and photometry was performed for the cases where we found
probable host galaxy
candidates. 

\begin{figure*}
\epsscale{1.82}
\vspace{-2cm}
\plotone{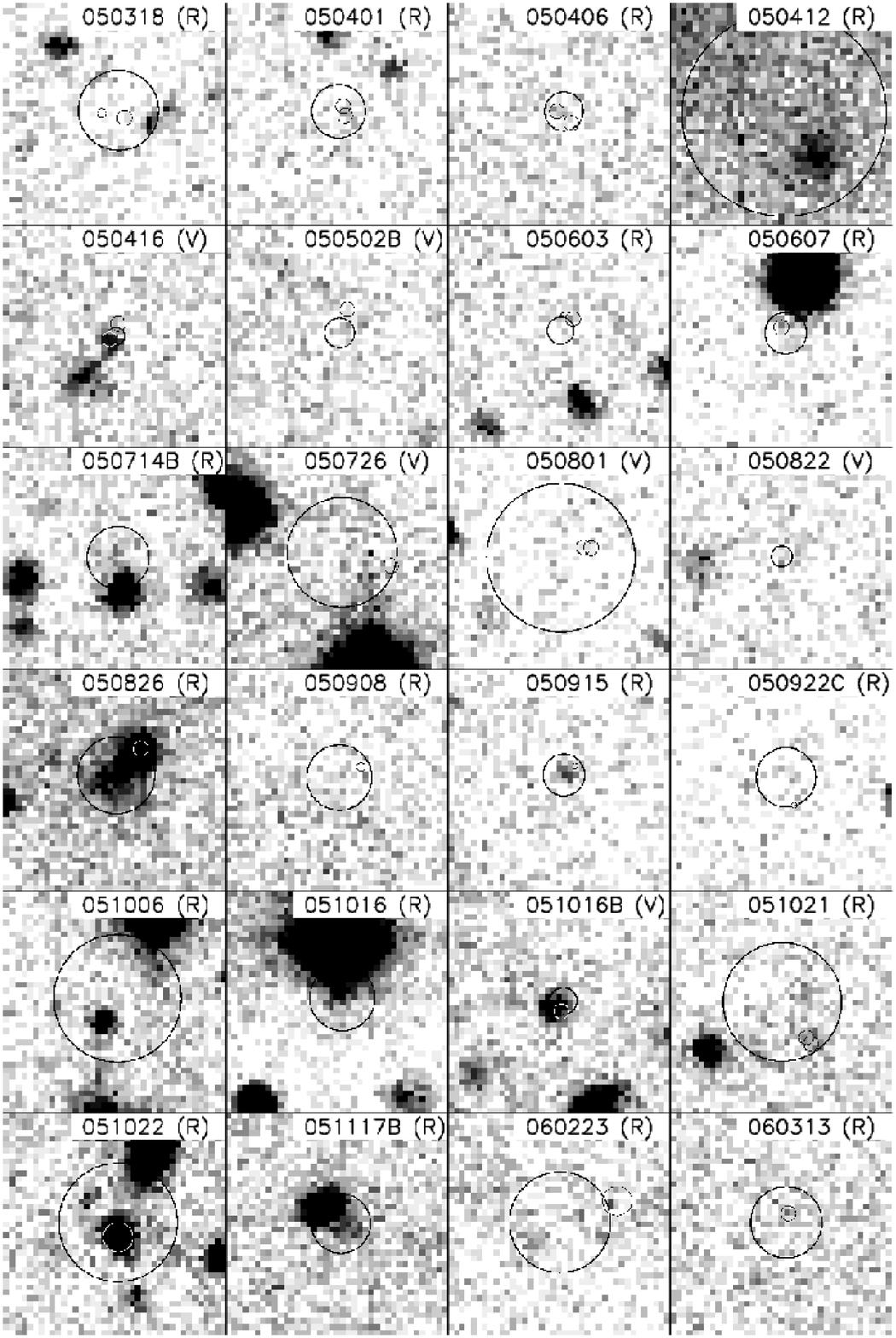}
\caption{Our GRB sample. Thick, large circles are the (refined) {\it Swift}
    XRT error circle and any smaller circles correspond to reported
    (optical, IR or radio) afterglow positions, see the text and
    Table~\ref{t:grb_info}. Each image is $15\arcsec \times
    15\arcsec$, with greyscale cuts of ${}^{+5}_{-0.5}\sigma_B$. The
    seeing, exposure time and 2$\sigma$-detection limit for each GRB
    field are found in Table~\ref{t:obs}. Several OT positions are
    lacking error estimates, see Table~\ref{t:grb_info}; in these cases we plot
    a 0$\farcs$5 error radius. For the figures, the original frames
    were rotated and resampled in order to make North up and East to the left.
    \label{f:grbs}}
\end{figure*}

Each object in our GRB sample is treated below; please consult
Fig.~\ref{f:grbs}, Table~\ref{t:grb_info} and Table~\ref{t:obs} for
details about the target GRBs, the data (filter, exposure time, and
seeing) and for $2\sigma$ detection limits, photometry and positions
of host galaxy candidates. The upper limits and host galaxy magnitudes
in Table~\ref{t:obs} are not corrected for foreground Galactic
extinction. The reader will also find that the detection limit (column
5) typically is different for images with roughly the same exposure
time and seeing; this is mainly due to differences in moon
illumination and sky brightness.  Quite often positions of afterglow
candidates in GCN\footnote{Gamma-ray burst Coordinates Network}
circulars are lacking error estimates, and in such cases we adopt a
0$\farcs$5 error radius in the figures. Also, we only include optical
afterglow positions which have been reported to be \emph{transient}.
The tables, the mosaic figure (from upper left to lower right) and the
following subsections are all arranged chronologically.  Throughout
this paper we use a ($\Omega_\mathrm{m}$,$\Omega_\Lambda$) = (0.3,0.7)
cosmology with Hubble Constant $H_0=70$ km s$^{-1}$ Mpc$^{-1}$.

\begin{deluxetable}{lllllllll}
\tabletypesize{\scriptsize} 
\tablecaption{GRB target list \label{t:grb_info}}
\tablewidth{0pt}
\tablecolumns{8}
\tablehead{
\multicolumn{1}{l}{GRB} & \multicolumn{1}{l}{Redshift} &
\multicolumn{3}{c}{XRT position (J2000)\tablenotemark{a}} &
\multicolumn{4}{c}{OT position (J2000)} \\
 &  & \colhead{RA} & \colhead{Dec} & \colhead{Error} &
\colhead{RA} & \colhead{Dec} & \colhead{Error} & \colhead{Ref.}
}
\startdata
050318  & 1.44 $[1]$ &  03:18:51.04 & $-$46:23:43.5 & 2.7 &
03:18:51.0  & $-$46:23:44 & 0.5 & $[11]$ \\
 & & & & & 03:18:51.15 & $-$46:23:43.7 & 0.3 & $[12]$ \\
%
050401 & 2.90 $[2]$ &  16:31:28.84 & $+$02:11:14.5 & 1.8 &
16:31:28.82  & $+$02:11:14.83 & 0.5 & $[13]$ \\
 & & & & & 16:31:28.81 & $+$02:11:14.2 & \nodata & $[14]$ \\
%
050406 & \nodata &  02:17:52.25 & $-$50:11:15.0 & 1.3 &
02:17:52.3 & $-$50:11:15 & 0.5 & $[15]$ \\
 & & & & & 02:17:52.2 & $-$50:11:15.8 & \nodata & $[16]$ \\
%
050412 & \nodata &  12:04:25.18 & $-$01:12:00.8 & 6.9 & \nodata &
\nodata &  \nodata \\
%
050416 & 0.65 $[3]$ &  12:33:54.57 & $+$21:03:26.9 & 0.6 &
12:33:54.6 & $+$21:03:26.7 & \nodata & $[17]$ \\
 & & & & & 12:33:54.56 & $+$21:03:27.73 & \nodata & $[18]$ \\
%
050502B & \nodata &  09:30:10.06 & $+$16:59:46.5 & 1.0 &
09:30:10.024 & $+$16:59:48.07 & \nodata & $[19]$ \\
%
050603  & 2.82 $[4]$ &  02:39:56.90 & $-$25:10:55.7 & 0.9 &
02:39:56.891 & $-$25:10:54.6 & 0.1 & $[20]$ \\
 & & & & & 02:39:56.839 & $-$25:10:54.92 & \nodata & $[21]$ \\
%
050607  & \nodata &  20:00:42.77 & $+$09:08:31.1 & 1.4 &
20:00:42.79 & $+$09:08:31.5 & 0.5 & $[22]$ \\
%
050714B & \nodata &  11:18:47.75 & $-$15:32:49.3 & 2.1 & \nodata &
\nodata & \nodata \\
%
050726 &  \nodata & 13:20:12.16 & $-$32:03:51.0 & 3.7 &
13:20:11.9 & $-$32:03:51.9 & \nodata & $[23]$ \\
%
050801 & \nodata & 13:36:35.51 & $-$21:55:42.7 & 5.0 &
13:36:35.4 & $-$21:55:42.0 & \nodata & $[24]$ \\
 & & & & & 13:36:35.363 & $-$21:55:42.03 & \nodata & $[25]$ \\
%
050822 & \nodata &  03:24:27.22 & $-$46:02:00.0 & 0.7 & \nodata &
\nodata & \nodata \\
%
050826 & 0.30 $[5]$ &  05:51:01.69 & $-$02:38:37.6 & 2.6 &
05:51:01.58 & $-$02:38:35.8 & 0.5 & $[26]$ \\
%
050908 & 3.34 $[6]$ &  01:21:50.85 & $-$12:57:17.9 & 2.2 &
01:21:50.75 & $-$12:57:17.2 & 0.3 & $[27]$ \\
%
050915 & \nodata &  05:26:44.86 & $-$28:00:59.9 & 1.4 &
05:26:44.804 & $-$28:00:59.27 & 0.18 & $[28]$ \\
%
050922C & 2.20 $[7]$ &  21:09:33.12 & $-$08:45:28.3 & 2.0 &
21:09:33.083 & $-$08:45:30.2 & 0.2 & $[29]$ \\
%
051006 & \nodata &  07:23:14.03 & $+$09:30:21.9 & 4.3 & \nodata &
\nodata &  \nodata \\
%
051016 & \nodata &  08:11:16.77 & $-$18:17:53.7 & 2.2 & \nodata &
\nodata & \nodata \\
%
051016B & 0.94 $[8]$ &  08:48:27.80 & $+$13:39:20.7 & 0.9 &
08:48:27.81 & $+$13:39:20.0 & \nodata & $[30]$ \\
%
051021\tablenotemark{\dag} & \nodata & 
01:56:36.5 & $+$09:04:06.1  &4.0 & 
01:56:36.37 & $+$09:04:03.27 & 0.5 & $[31]$ \\
 & & & & & 01:56:36.39 & $+$09:04:03.7 & 0.5 & $[32]$ \\
%
051022\tablenotemark{\dag} & 0.8\phm{0} $[9]$ & 
23:56:04.1 & $+$19:36:25.1 & 4.0 & 
23:56:04.1 & $+$19:36:24.1 & 1.0 & $[33]$\tablenotemark{\ddag} \\
%
051117B & \nodata &  05:40:43.21 & $-$19:16:27.2 & 2.0 & \nodata &
\nodata & \nodata \\
%
060223 & 4.41 $[10]$ &  03:40:49.82 & $-$17:07:49.8 & 3.4 &
03:40:49.55 & $-$17:07:48.36 & 1.0 & $[34]$ \\
%
060313 & \nodata &  04:26:28.41 & $-$10:50:40.7 & 2.4 &
04:26:28.4 & $-$10:50:40.1 & 0.5 & $[35]$ \\
%
\enddata
\tablecomments{XRT and OT positions (with error radii in arcsec) for
  the GRBs in our sample, see also Fig.~\ref{f:grbs}.
References:
$[1]$ \citet{GCN3122}, 
$[2]$ \citet{GCN3176}, 
$[3]$ \citet{GCN3542}, 
$[4]$ \citet{GCN3520}, 
$[5]$ \citet{GCN5982},
$[6]$ \citet{GCN3971}, 
$[7]$ \citet{GCN4029}, 
$[8]$ \citet{GCN4186}, 
$[9]$ \citet{GCN4156}, 
$[10]$ \citet{GCN4815}, 
$[11]$ \citet{GCN3114}, 
$[12]$ \citet{GCN3123}, 
$[13]$ \citet{GCN3187}, 
$[14]$ \citet{GCN3163}, 
$[15]$ \citet{GCN3185}, 
$[16]$ \citet{GCN3186}, 
$[17]$ \citet{GCN3265}, 
$[18]$ \citet{GCN3276}, 
$[19]$ \citet{GCN3338}, 
$[20]$ \citet{GCN3513}, 
$[21]$ \citet{GCN3516}, 
$[22]$ \citet{GCN3527}, 
$[23]$ \citet{GCN3698}, 
$[24]$ \citet{GCN3723}, 
$[25]$ \citet{GCN3736}, 
$[26]$ \citet{GCN4749}, 
$[27]$ \citet{GCN3945}, 
$[28]$ \citet{GCN3990}, 
$[29]$ \citet{GCN4015}, 
$[30]$ \citet{GCN4105}, 
$[31]$ \citet{GCN4159}, 
$[32]$ \citet{GCN4120}, 
$[33]$ \citet{GCN4154}, 
$[34]$ \citet{GCN4813}, 
$[35]$ \citet{GCN4871}.
}
\tablenotetext{a}{All XRT positions are from \citet{Butler06}, except
  for the two HETE bursts 051021 and 051022, see text.
} \tablenotetext{\dag}{{\it HETE}-localized burst (the rest are {\it
    Swift} bursts)} \tablenotetext{\ddag}{Radio transient}
\end{deluxetable}

\begin{deluxetable}{lcrllllc}
\tabletypesize{\scriptsize} 
\tablecaption{Data set and photometry \label{t:obs}}
\tablewidth{0pt}
\tablehead{
\colhead{GRB} & \colhead{Band\tablenotemark{a}} &
\colhead{Exp.~time\tablenotemark{b}} & \colhead{Seeing\tablenotemark{c}} &
\colhead{2$\sigma$-limit\tablenotemark{d}} & \colhead{Host candidate} &
\colhead{RA \& Dec (J2000)\tablenotemark{e}} & \colhead{OT\tablenotemark{f}}
}
\startdata
050318 & $R$& 36\,000 & 1.0 & 26.0 & \nodata & \nodata &  y \\
\tableline
050401 & $R$ &  7\,200 & 1.2  & 25.1 & \nodata & \nodata &  y \\
          & $V$ & 14\,000 & 1.15 & 25.7 & \nodata & \nodata \\
\tableline
050406 & $R$ & 8\,100 & 1.2 & 25.1 & \nodata & \nodata & y \\
\tableline
050412 &$R$ & 5\,400& 1.75 & 24.6 & 22.4 (0.2) &12:04:25.03 $-$01:12:04.0& n \\
          & $I$ & 3\,000 & 0.8  & 24.3 & 21.6 (0.1) & \nodata \\
          & $V$ & 8\,250 & 1.05 & 25.2 & 23.4 (0.2) & \nodata \\
\tableline
050416 & $V$& 15\,900 & 1.3 & 25.8 & 24.2 (0.2) &12:33:54.59 $+$21:03:26.6& y\\
\tableline
050502B & $V$ & 31\,400 & 1.1 & 26.3 & \nodata & \nodata & y \\
\tableline
050603  & $R$ & 17\,400 & 1.15 & 25.4 & \nodata & \nodata & y \\
         & $I$ & 7\,200  & 1.0 & 24.4 & \nodata & \nodata \\
\tableline
050607  & $R$ & 1\,980  & 1.1  & 24.0 & \nodata & \nodata & y \\
           & $I$ & 2\,400  & 1.55 & 22.6 & \nodata & \nodata \\
\tableline
050714B & $R$ & 15\,000 & 1.15 & 25.8 & 21.8 (0.1) &11:18:47.72 
                                              $-$15:32:51.6 & n \\
           & $I$ & 8\,700  & 0.95 & 24.3 & 20.6 (0.1) & \nodata \\
           & $V$ & 27\,400 & 1.05 & 26.3 & 22.9 (0.1) & \nodata \\
           & $B$ & 15\,600 & 1.1  & 26.1 & 24.9 (0.3) & \nodata \\
\tableline
050726 & $V$ & 15\,900 & 1.0 & 25.9 & \nodata & \nodata & y \\
\tableline
050801 & $I$ & 65\,550 & 1.15 & 24.8 & \nodata & \nodata & y \\
          & $V$ & 22\,500 & 1.1  & 26.2 & \nodata & \nodata \\
          & $B$ & 24\,000 & 1.05 & 26.3 & \nodata & \nodata \\
\tableline
050822 & $V$ & 9\,600 & 1.7 & 25.0 & \nodata & \nodata & n \\
\tableline
050826 & $R$& 27\,000 & 0.95 & 25.8 & 21.4 (0.1) & 
 05:51:01.59 $-$02:38:36.1 & y \\
          & $I$ &  3\,000 & 1.0 & 23.8 & 20.6 (0.2) & \nodata \\
           & $V$ &  6\,000 & 1.1  & 25.6 & 22.4 (0.2) & \nodata \\
           & $B$ &  5\,400 & 1.1  & 25.4 & 23.9 (0.3) & \nodata \\
\tableline
050908  & $R$ & 3\,000 & 0.75 & 25.2 &  \nodata & \nodata & y \\
\tableline
050915& $R$& 22\,000 &0.95& 25.9 & 24.8 (0.4) & 
 05:26:44.84 $-$28:00:59.7 & y \\
          & $I$ & 10\,800 & 1.1 & 24.3 & \nodata   & \nodata \\
          & $V$ & 13\,800 & 1.0 & 25.9 & 25.2 (0.5) & \nodata \\
          & $B$ &  3\,000 & 1.1 & 25.5 & \nodata   & \nodata \\
\tableline
050922C & $R$ & 23\,700 & 0.95 & 25.8 & \nodata & \nodata & y \\
\tableline
051006 &$R$ & 9\,600 & 0.95 & 25.6 & 23.0 (0.1) &07:23:14.10 $+$09:30:20.2& n\\
          & $I$ & 5\,400 & 0.9  & 24.3 &  22.2 (0.2) &  \nodata \\
          & $V$ & 7\,200 & 1.1  & 25.6 &  23.3 (0.2) &  \nodata \\
          & $B$ & 6\,600 & 1.1  & 25.6 &  24.2 (0.2) &  \nodata \\
\tableline
051016 & $R$ & 24\,900 & 0.95 & 25.6 & \nodata & \nodata & n \\
\tableline
051016B & $V$& 16\,800 & 1.05 & 26.1 & 23.1 (0.2) &08:48:27.84 $+$13:39:20.2
 & y \\
\tableline
051021\tablenotemark{\dag}
 & $R$ & 19\,800 & 1.1 & 25.4 & 24.9 (0.4) & 
 01:56:36.35 +09:04:03.7  & y \\
           & $I$ & 11\,400 & 1.2 & 24.0 &   \nodata & \nodata \\
           & $V$ &  6\,000 & 1.3 & 25.2 &   \nodata & \nodata \\
           & $B$ &  1\,800 & 1.25 & 24.6 &  \nodata & \nodata  \\
\tableline
051022\tablenotemark{\dag}
 & $R$& 15\,000 & 1.0  & 25.4 & 21.7 (0.1) & 
 23:56:04.10 +19:36:24.2 & n \\
           & $I$ & 13\,800 & 1.3  & 23.8 & 21.1 (0.1) & \nodata \\
           & $V$ & 12\,600 & 1.15 & 25.7 & 22.3 (0.1) & \nodata \\
           & $B$ & 12\,600 & 1.2  & 25.7 & 22.8 (0.2) & \nodata \\
\tableline
051117B&$R$& 10\,200 &1.1& 25.1 & 21.0 (0.1) & 
 05:40:43.29 $-$19:16:26.1 & n \\
           & $I$ & 13\,650 & 1.15 & 24.0 & 20.4 (0.1) & \nodata \\
           & $V$ &  9\,000 & 1.0 & 25.2 & 21.8 (0.2) &  \nodata \\
           & $B$ & 12\,600 & 1.0 & 25.3 & 22.5 (0.2) &  \nodata \\
\tableline
060223 & $R$ & 14\,600 & 1.2 & 25.3 & \nodata & \nodata & y \\
\tableline
%
%
060313 & $R$ & 8\,100 & 1.30 & 23.9 & \nodata & \nodata & y \\
\enddata
\tablecomments{Details regarding the data set and our photometric results. The
  host galaxy candidates for GRBs 050915 and 051021 are new and have
  not been suggested or published by others.}
\tablenotetext{a}{Calibrated to the Landolt $BVR_CI_C$ system.}
\tablenotetext{b}{Total exposure time (seconds) of stacked image.}
\tablenotetext{c}{FWHM seeing (arcseconds) of stacked image.}
\tablenotetext{d}{See definition of detection limit in Sect.~\ref{S:phot}.} 
\tablenotetext{e}{Estimated errors are 0$\farcs$5 in both
  coordinates.}
\tablenotetext{f}{Optical transient (yes or no), see also
  Table~\ref{t:grb_info}.}
\tablenotetext{\dag}{{\it HETE}-localized burst}
\end{deluxetable}
\clearpage

\subsection{GRB 050318} 
Sixty 600 s exposures, some under bright sky conditions, yield a
detection limit of $R=26.0$ for this $z=1.44$ burst \citep{GCN3111,
  GCN3122}. The optical afterglow positions reported by
\citet[][{\it Swift}-UVOT]{GCN3123} and \citet{GCN3114} are not fully
consistent, however both are inside the refined XRT error circle of
\citet{Butler06}.

We do not find any sources in immediate proximity to the {\it
  Swift}-UVOT error circle of \citeauthor{GCN3123}, but an extended
source is seen at the border of the XRT error circle, less than
2$\arcsec$ west of the \citeauthor{GCN3114} position
(see Fig.~\ref{f:grbs}). The light distribution has two peaks a little
more than 1$\arcsec$ apart, so it could potentially be two separate
(possibly interacting) galaxies or two bright spots (e.g.\
star-forming regions) in the same galaxy.  Using an elliptical
aperture around the entire light distribution, we estimate the
magnitude to be $R=23.8 \pm 0.2$. However, we err on the side of
conservatism and exclude this source as a host candidate, since OTs
typically are highly concentrated on the very brightest regions of
their host galaxies \citep[e.g.][]{Fruchter06}. At $z=1.44$ the
separation between the source (maximum pixel) and the OT is about 16
kpc, which is considered too large in terms of long bursts whose
progenitors are assumed to be massive stars. If the error bar on the
reported OT position of \citet{GCN3114} was larger, the object found
here could qualify as a host candidate.

\subsection{GRB 050401} 
GRB 050401 \citep{GCN3161,dePasquale06} is a high-redshift burst;
\citet{GCN3176} obtained spectra of the afterglow and found $z=2.90$
from several absorption lines \citep[see also][]{Watson06}. Optical
and radio afterglow positions are reported by \citet{GCN3163} and
\citet{GCN3187}, respectively.  Our limits of $R=25.1$ and $V=25.7$
are not deep enough to reveal any host galaxy.

\subsection{GRB 050406} 
Two afterglow positions \citep{GCN3185, GCN3186} were reported for GRB
050406 \citep{GCN3180}, and both are shown in Fig.~\ref{f:grbs}. No
sources are found in or near the XRT error circle down to the limiting
magnitude of $R=25.1$. 

\subsection{GRB 050412} 
The field around GRB 050412 \citep{GCN3237} is contaminated by a
bright $R=11.3$ star (USNO U0825-07638162) approximately $50\arcsec$
north-east. No transient source has been confirmed in optical bands,
although \citet{GCN3243} and \citet{GCN3244} detected a single source
and quoted approximate values of $R \sim 22$ (1.5~h after the burst)
and $R \gtrsim 21.5$ (55 min after the burst), respectively.  Later,
\citet{GCN3263} found another source inside the XRT error circle
estimated at $R=26.0 \pm 0.5$, but was unable to determine whether it
was point-like or extended.

We find a clearly extended object inside the XRT error
circle. $VRI$-photometry and position are listed
in Table~\ref{t:obs}. In particular, our measurement of $R=22.4\pm
0.2$ are in rough agreement with \citet{GCN3243} and \citet{GCN3244},
considering their very preliminary and approximate values, and the
positions are also coincident. This suggests that the source reported
by the above authors was not the OT. We regard this extended object as
a possible host galaxy.

Unfortunately, our images are not deep enough to probe the faint
$R=26.0 \pm 0.5$ detection by \citet{GCN3263}. Further imaging to
search for this object, as well as spectroscopy of the host candidate
presented here, are advised.

\subsection{GRB 050416} 
An optical afterglow for GRB 050416 \citep{GCN3264} was first found by
\citet{GCN3265} and later confirmed by \citet{GCN3276} and other groups,
also at IR and radio wavelengths. \citet{GCN3542} used Keck I spectra
to calculate a redshift of $z=0.6535$ from the host galaxy's emission
lines. 
See also \citet{Soderberg06} and \citet{Holland07}.

Our $V$-band image clearly detects the galaxy (along with another
extended source outside the XRT error circle), see
Fig.~\ref{f:grbs}. The magnitude is estimated to be $V=24.2 \pm 0.2$
and the position is given in Table~\ref{t:obs}.

\subsection{GRB 050502B} 
We only carried out $V$-band imaging for GRB 050502B \citep{GCN3330,
  Falcone06} and reached a limit of 26.3 mag. The stacked image shows
no sign of any host near the \citet{GCN3338} candidate afterglow
position (transient was only seen in the $I$-band) or inside the XRT
circle.

\subsection{GRB 050603} 
GRB 050603 \citep{GCN3509,Grupe06} have both optical \citep{GCN3511,GCN3516}
and radio \citep{GCN3513} afterglows, but no imaging of the host
galaxy. 2.13 days after the burst \citet{GCN3520} found a redshift of
$z=2.821$ based on a single bright emission line interpreted as
Ly$\alpha$.

Our $R$- and $I$-band images, with limits 25.4 mag and 24.4 mag,
respectively, do not show any sources near the afterglow positions.

\subsection{GRB 050607} 
The host of GRB 050607 \citep{GCN3525,Pagani06b} will be hard to study due to
the nearby $R=16.6$ USNO star U0975-17511046, see Fig.~\ref{f:grbs}.
Our rather shallow limits of $R=24.0$ and $I=22.6$ do not reveal any
other objects near the afterglow position of \citet{GCN3527}.
PSF-subtraction is difficult as the star is saturated in our images.

\subsection{GRB 050714B} 
GRB 050714B \citep{GCN3613} had an initial XRT error circle where four
sources reportedly where candidates for either the OT or host galaxy.
The refined XRT position of \citet{Butler06} only includes one of
them, i.e.\ source \#1 from \citet[][see also
\citealp{GCN3614}]{GCN3616}, which was found to be non-variable with
an estimated $R=21.7$ mag. None of the reported sources have been
confirmed to fade and are thus not considered as OT candidates.
Additionally, we detect all of them in our images taken months after
the burst.

Down to the limits given in Table~\ref{t:obs}, we only detect the
object mentioned above (partly) inside the refined XRT error circle.
We consider it a host candidate, although $BVRI$ photometry yields a
very red color and it is unresolved at the resolution of the images.
Our $R$-band magnitude of $21.8\pm 0.1$ agrees with \citet{GCN3616}.


\subsection{GRB 050726} 
\citet{GCN3698} reported a detection for this burst \citep{GCN3682} by
{\it Swift}-UVOT only in the $V$-band, with $V=17.35 \pm 0.09$ at
123~s after the burst, and then dropping below the detection
threshold.  This detection has not been confirmed by others.

South-east of the {\it Swift}-UVOT position our $V$-band image (taken
months after the burst) shows a few pixels with values above the
background level, see Fig.~\ref{f:grbs}. However, at least some of
this signal is due to a diffraction spike from the nearby star.  Thus
is not possible to say whether this is a very weak detection of the host
galaxy. The limiting magnitude for this image is $V=25.9$.

\subsection{GRB 050801} 
The ROTSE collaboration reported the first optical afterglow
\citep{GCN3723,Rykoff06} of {\it Swift} trigger 148522. Later, \citet{GCN3736}
found an OT position consistent with \citet{GCN3723}, while the
position from {\it Swift}-UVOT \citep{GCN3733} is at variance with the
aforementioned positions and even outside the XRT error circle.
We do not detect any host candidates in the immediate surroundings,
despite rather deep limits as far as our observations are concerned,
see Table~\ref{t:obs} and Fig.~\ref{f:grbs}.

\subsection{GRB 050822} 
Our images of GRB 050822 \citep{GCN3849,Godet07} were taken in rather
poor seeing and yielded an upper limit of $V=25.0$.
With the exception of the X-ray band, there have been no reports
concerning transient objects.

\subsection{GRB 050826} \label{s:grb050826} 
\citet{GCN3891} claim the detection of a fading optical afterglow for
GRB 050826 \citep{GCN3884} located at the border of the refined XRT
position, see Fig.~\ref{f:grbs}. From August 26 to 27
\citeauthor{GCN3891} measure a decrease in brightness from $R=21.0$ to
$R=21.6$ (both $\pm0.2$ mag). In a later GCN, \citet{GCN4749} provide
positions for the presumed host galaxy and OT candidate, and recently
the redshift of the host candidate was measured to be $z=0.297$
\citep{GCN5982}.

In our 27\,000~s stacked $R$-image we find an extended source with
magnitude $R=21.4 \pm 0.1$. The source seems to consist of a
conspicuously point-like object (position: see Table~\ref{t:obs}) plus
an extended, irregular and much fainter source in terms of maximum
pixel intensity. The resolved part lies in the very center of the
refined XRT error circle. The quoted $BVRI$-magnitudes in the table
were found using elliptical apertures covering the \emph{entire} light
distribution. We note that our position for the point-like source is
consistent with both the OT and host position of \citet{GCN4749}.

This field is the only one in our sample with a significant foreground
Galactic extinction: $E_{B-V}=0.6$ according to the dust maps of
\citet{Schlegel98}. 



\subsection{GRB 050908} 
The observations of GRB 050908 \citep{GCN3942} comprise only five 600
s exposures, but favorable seeing yields a detection limit of
$R=25.2$.  OT positions are provided by \citet{GCN3943} and
\citet{GCN3945}; the latter one is plotted in Fig.~\ref{f:grbs}. Given
a burst redshift of $z=3.344$ \citep{GCN3971}, it is not surprising
that the host galaxy is undetected in our image.

\subsection{GRB 050915} 
An IR counterpart \citep[$H$-band,][]{GCN3990} has been observed for
GRB 050915 \citep{GCN3977}, but no redshift or host galaxy has been
found. However, close to the IR afterglow position by
\citeauthor{GCN3990} both our $R$- and $V$-band images reveal 7 pixels
above the 2$\sigma$ noise level, see Fig.~\ref{f:grbs}.
Images in filters $I$ and $B$ have limiting magnitudes of 24.3 and 25.5,
respectively, and yield no signal.

Aperture photometry is highly uncertain for such a marginal detection.
However, if we assume that the light distribution of the host
candidate follows the PSF, one can construct a synthetic model and
resample it on the pixel array nearby the candidate. The strength of
the synthetic source is increased until the detected light is similar
to the host candidate. This gives $R = 24.8 \pm 0.4$ and $V = 25.2 \pm
0.5$. The errors are obviously rather large and this method rests on
the assumption that the brightness profile of the detected part of the
host is similar to the image PSF. However, a distant and/or faint
galaxy might easily fulfill this assumption.



\subsection{GRB 050922C} 
For GRB 050922C \citep{GCN4013} we have a total of 23\,700 s in the
$R$-band, giving a detection limit of $R=25.8$. The host at $z=2.198$
\citep[][redshift based on absorption features in the optical
afterglow]{GCN4029} must be fainter than this limit as our
observations fail to detect anything inside the XRT error circle and,
more specifically, at the well-constrained optical afterglow position
of \citet{GCN4015}.

\subsection{GRB 051006} 
Inside the XRT error circle of GRB 051006
\citep{GCN4061} we find in our images a point-like source.
This is the same object as the ``\#2 source'' mentioned by
\citet{GCN4064}, who was not able to check for variability and did not
estimate the brightness. (Regarding the other objects in early
reports: source \#1 is a USNO star and source \#3 is about 5$\arcsec$
outside the refined XRT error circle.) In the absence of a confirmed
optical/IR/radio afterglow, it seems that the object mentioned by
\citet{GCN4064} was an existing source, for which we here present
multi-color photometry (Table~\ref{t:obs}). We consider it a host
candidate. It is the only detected object inside the XRT error circle.


\subsection{GRB 051016} 
We do not detect any sources inside the XRT error circle of GRB 051016
\citep{GCN4096} down to a limiting magnitude of $R=25.6$. No OTs are
reported. As seen in the figure, a $R=16.7$ star (USNO U0675-08481584)
contaminates the area, and PSF-subtraction of this partly saturated
star reveals no apparent host candidates.


\subsection{GRB 051016B} 
\citet{GCN4186} found a redshift of $z=0.9364$ for GRB 051016B
\citep{GCN4103} after observing the OT position by \citet{GCN4105}
about 15 days after the burst. The redshift determination was based
on two emission lines ([O~{\small II}] and Ne~{\small III}) in spectra
from the Keck~I telescope.

We here present the first image of the host galaxy, which is
unresolved in our images, see Fig.~\ref{f:grbs}. It is located near
the OT position of \citeauthor{GCN4105}, and we measure $V=23.1 \pm
0.2$ mag.

\subsection{GRB 051021} 
\begin{figure}[ht]
\epsscale{1}
\plotone{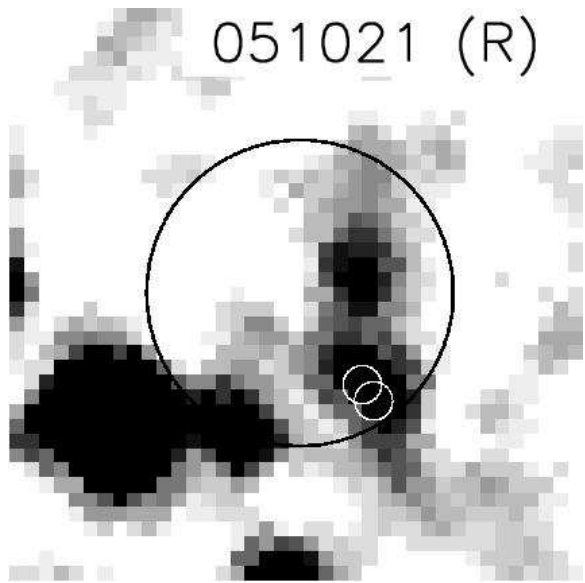} 
\caption{Smoothed $R$-band image of GRB 051021. The two small circles
  correspond to the OT positions mentioned in the text and in
  Table~\ref{t:grb_info}.
  \label{f:051021_smooth}}
\end{figure}

This {\it HETE} burst \citep{GCN4116} has no reported redshift or host
galaxy. At the position of the optical afterglow
\citep{GCN4120,GCN4159} we detect 4--5 neighboring pixels with values
at the 2$\sigma$ noise level in our stacked $R$-band image, see
Fig.~\ref{f:grbs}. Formally, this detection is of low significance,
but combining all data from all four bands strengthens the presence of
an excess flux on this area. Moreover, Gaussian smoothing of the
$R$-band image reveals a faint, but significant source just at the OT
position, see Fig.~\ref{f:051021_smooth}. We estimate $R=24.9 \pm 0.4$
for this likely host galaxy of GRB 051021.

\subsection{GRB 051022} 
Despite extensive observational efforts from many groups no optical
afterglow was found for this {\it HETE} burst \citep{GCN4131,Nakagawa06}.
However, the host galaxy was pinpointed and a redshift of $z=0.8$ was
deduced from preliminary reductions where a strong line was
interpreted as O{\small II} \citep{GCN4156}. \citet{GCN4154} also
found a candidate radio afterglow coincident with the host galaxy (see
Fig.~\ref{f:grbs}). In Table~\ref{t:obs} we present 
($BVRI$) photometry for the host galaxy of GRB 051022.



\subsection{GRB 051117B} 
GRB 051117B \citep{GCN4281} has no known optical counterpart or
redshift. A possible host was suggested by \citet[][see also
\citealt{GCN4285}]{GCN4291} using part of the data presented in this
article.

In Table~\ref{t:obs} we present $BVRI$ photometry and a refined
position for this host candidate. We do not find any other objects
near the XRT error circle down to the limiting magnitudes in
Table~\ref{t:obs}.

\subsection{GRB 060223} 
\citet{GCN4815} found a redshift of $z=4.41$ for GRB 060223
\citep{GCN4813}, and only the reddest band ($V$) of {\it Swift}-UVOT
detected the OT of this high-$z$ event.

Two neighboring pixels inside the OT error circle \citep{GCN4813} are
at the 3$\sigma$ level above the background (see Fig.~\ref{f:grbs}),
but we can obviously not claim detection of a source. We note that our
limiting magnitude for this GRB field is $R=25.3$.

\subsection{GRB 060313}  
This is the only short burst in our sample \citep{GCN4867,Roming06}.
The first afterglow report was by \citet{GCN4871}.
The first night's images contained the OT, so later observations were
used to search for the host galaxy. Due to unfavorable seeing and moon
illumination our $R$-band limit is only 23.9 mag, and no sources are
seen in the XRT error circle.


\section{Comparison with pre-\emph{Swift} GRB host sample}
\label{S:comparison}

\begin{figure}[!htb]
\plotone{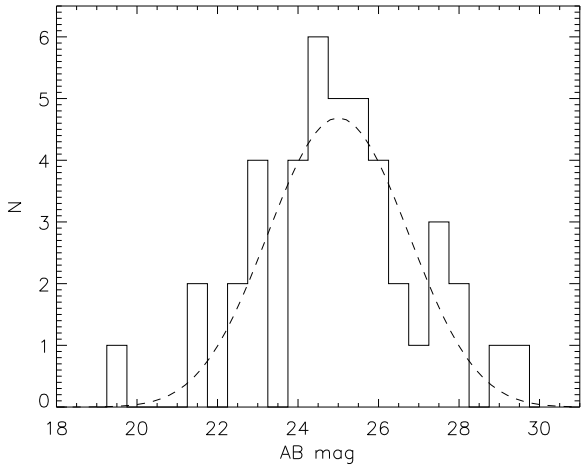} 
\caption{Histogram of the pre-{\it Swift} GRB host galaxy sample of
  \citet{Fruchter06}. Bin size = 0.5 mag. The dotted line is a beta
  function fit to the data (see text).\label{f:Fruchter_hist}}
\end{figure}

\citet{Fruchter06} have compiled a sample of 46 galaxies that hosted
long GRBs with well-localized OT
positions 
detected prior to the launch of the {\it Swift} satellite.  Discarding
two galaxies without a brightness estimate and the somewhat
controversial and nearby $z=0.0085$ GRB 980425, this pre-{\it Swift}
sample contains photometry in the AB photometric system for 43 GRB
host galaxies -- see Fig.~\ref{f:Fruchter_hist}. The quoted AB
magnitudes are a mix of several different filters, primarily {\it HST}
``$V$-like'' filters -- F606W, F555W, and ``CLEAR'' -- but also F775W
(two cases) and F814W (one case).
Our sample, i.e.\ the magnitudes from Table~\ref{t:obs}, can not be
\emph{directly} compared with the \citeauthor{Fruchter06} sample for
several reasons: First, our sample contains several bursts without a
confirmed OT. Second, our table lists $R$- and $V$-magnitudes, not AB
magnitudes, for the hosts and upper limits. Third, we detected a host
galaxy candidate only for a few bursts. This last point renders
several standard statistical methods for sample comparison (e.g.\ a
direct Kolmogorov-Smirnov test) inapplicable.  Instead we have applied
the upper limits in the analysis and tested, using Monte Carlo
simulations, how many GRB hosts we should have detected based on the
distribution of host brightnesses from \citet{Fruchter06}.
At this point we want to note that the following comparison between
the two samples is not statistically robust, as neither our sample nor
the pre-{\it Swift} sample of \citeauthor{Fruchter06} is complete. If
we consider only long {\it Swift} GRBs with confirmed OTs, our data
include 50\% of all targets observable from La Silla during our two
observing runs.  \citeauthor{Fruchter06}'s sample (constrained by
having OTs and being in the long burst category) is probably less
complete, since only about a third of the detected pre-{\it Swift}
GRBs are sufficiently well localized to permit host identifications
\citep[e.g.,][]{Fynbo01,Berger02}.
We nevertheless choose to include this comparison as an exercise in
order to check whether our observations are comparable with the host
magnitude distribution in \citet{Fruchter06}

Before we carried out the simulations we first defined a subset of
the Danish 1.54m observations which comprises 15 \emph{long} GRBs
detected with {\it Swift} and which have a \emph{confirmed optical
  transient}: 050318, 050401, 050406, 050416, 050502B, 050603, 050607,
050726, 050801, 050826, 050908, 050915, 050922C, 051016B, and 060223.
Short bursts, {\it HETE} bursts, and bursts without a confirmed OT
were omitted.  Next, since the \citeauthor{Fruchter06} sample quotes
magnitudes in the AB system we transformed our $R$ and $V$
upper limits for the host galaxies into the AB system according to
\citet{Fukugita95}. Finally, dust maps from \citet{Schlegel98} were
used to correct for foreground Galactic extinction. The AB magnitudes
of both our and \citeauthor{Fruchter06}'s sample encompass bands from
the visible to the red; this will widen the host magnitude
distributions compared to samples observed with a single filter, but
this fact should not seriously affect the conclusions we draw below.
%

The rationale for performing Monte Carlo simulations is that we want
to see how many host galaxies we would have detected had the
distribution of host magnitudes of our targets matched that of the
\citeauthor{Fruchter06} sample. This will be compared to the four host
detections we claim (concerning GRBs 050416, 050826, 050915, and
051016B).

We need to quantify the \citeauthor{Fruchter06} host magnitude distribution;
this was done by fitting a beta distribution
function\footnote{Unnormalized beta distribution: $f(x;\alpha,\beta) \sim
  x^{\alpha-1}(1-x)^{\beta-1}$, with parameters $\alpha,\beta > 0$}
(which allows for skewness) to the magnitude histogram, see the dotted
line in Fig.~\ref{f:Fruchter_hist}.
Even though the sample contains 43 hosts, the sampling is rather
sparse in the wings. We tried fitting different functions to the
histogram, but this did not significantly change the end results of
the analysis.

\begin{figure}[!htb]
\plotone{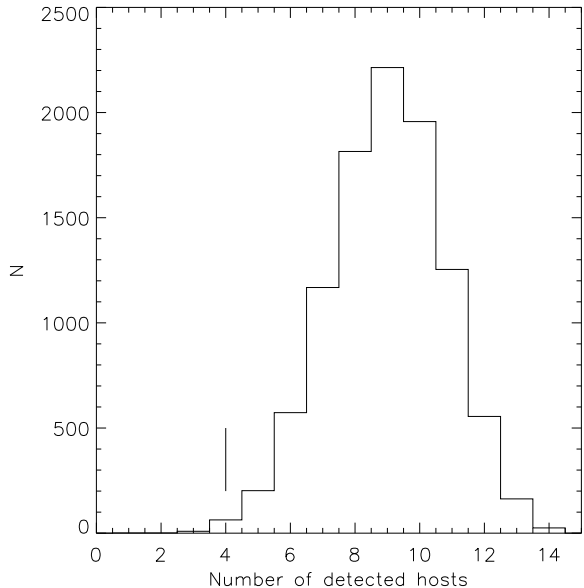} 
\caption{Histogram showing the distribution of host galaxy detections
  from the Monte Carlo simulations. We detected four hosts (indicated
  by the vertical line) from our 15 GRB fields, while the simulations
  indicate that we should have detected $9.0 \pm 1.8$ (1$\sigma$) had the
  host magnitude distribution of our sample been similar to the one in
  \citet{Fruchter06}.
  \label{f:MC}}
\end{figure}

Using the beta probability density function as an approximation of the
host magnitude distribution in \citeauthor{Fruchter06}, one can draw
magnitudes (using a random seed) and then compare them one by one to
our 15 upper limits. If the magnitude drawn is less -- i.e.\ brighter
-- than the upper limit (transformed to the AB photometric system and
corrected for foreground Galactic extinction, as described above) it
is recorded as a host detection. Having done this for all the 15 upper
limits one finally sums the number of detections. This procedure is
repeated $10\,000$ times, thus producing a distribution of the number
of host detections, see Fig.~\ref{f:MC}.
(We also checked that the simulated magnitudes based on the fitted
beta distribution was comparable to the original
\citeauthor{Fruchter06} sample.  The mean of the simulated values was
exactly the same as for the original sample, and the median and
the scatter was also similar.)

While we found \emph{four} host galaxies out of our 15 GRB subsample, the
mean number of detected host galaxies from the simulations is 9.0,
with a standard deviation of 1.8. In fact, 99.3\% of the Monte Carlo
simulations yielded five or more host detections.
This confidence level should not be taken at face value due to the
statistical incompleteness of the samples. This
coupled with the fact that the possible biases and selection
effects in each of them are hard to quantify, makes a definite and
quantitative conclusion impossible.
We have in this exercise simply tried to constrain our observed GRB
targets in the same way (solely long GRBs with confirmed OT) as in the
\citeauthor{Fruchter06} sample and then performed Monte Carlo runs to
investigate whether the two samples are comparable.
The above simulations indicate that the {\it Swift} GRB host sample
presented here is fainter than the pre-{\it Swift} host sample of
\citeauthor{Fruchter06} While this can not be extended to hosts in
general due to the incompleteness of both samples, we do consider it
tentative evidence of fainter {\it Swift}
hosts. 

Concerning possible biases, we want to make a couple of comments.
Since GRBs which were undetected in ground-based follow-ups often were
later chosen as targets for the HST programs on which the
\citeauthor{Fruchter06} data is based, one could argue that this would
bias the \citeauthor{Fruchter06} sample towards the faint end of the
pre-{\it Swift} host luminosity function. If so, this would strengthen
our findings.
Secondly, large Galactic extinctions can not explain or contribute to the
faintness of our GRB hosts, as only one field (i.e.\ GRB 050826, whose
host is in fact detected) has significant foreground extinction.

To summarize, we detect four hosts in a subsample of 15 GRBs selected
using comparable criteria to that of \citet{Fruchter06}. Assuming, for
a moment, that the host magnitude distribution of
\citeauthor{Fruchter06} is universal, we would expect to detect some
7--11 hosts given our observing conditions. If we had observed every
{\it Swift}-detected GRB field observable from La Silla during our two
runs we should have detected about twice as many hosts, i.e.\ about
18. To achieve this we would have to detect a host for just about
every remaining unobserved GRB field.  We can not find a plausible reason
why our observed sample should be selected in such a heavily biased
way. It seems reasonable that the most natural explanation for our
observed lower-than-expected rate of host detections is that the hosts
are fainter than the \citeauthor{Fruchter06} sample.

\section{Summary} \label{S:summary}

The large observational data set of 24 GRB fields obtained with the
Danish 1.54m telescope on La Silla have been used to search for the
galaxies hosting these bursts. New, previously unpublished, host
galaxy candidates are presented for GRBs 050915 and 051021. We also
suggest possible hosts for GRBs 050412, 050714B, 050826, 051006, and
051117B; these objects are reported in early GCN circulars as part of
the search for the OT, but since we detect them months after the
bursts they are not transient sources and, consequently, qualify as at
least candidate hosts.  Photometry for the host galaxies of GRBs
050416, 051016B, and 051022 is also provided.  These galaxies already
have measured redshifts from emission lines, but to our knowledge no
previously published photometry. Magnitudes (not corrected for
Galactic extinction) and positions are given in Table~\ref{t:obs}, as
well as upper limits in various photometric bands for the
non-detections.


We fail to detect more than half of the hosts in our GRB sample,
despite rather deep limits. These fields, along with a couple of our
marginal detections, e.g.\ GRBs 050915 and 051021, should be imaged
with larger telescopes.  Likewise, the new host candidates presented
here are prime targets for spectroscopic follow-up to determine the
redshift and probe extinction, star-formation rate etc. Redshift
determinations for the hosts would also fix the energy and time scale
of the bursts and afterglows, and thus contribute to the increasing
data base and knowledge on the GRB phenomenon.


It is worth noting that as far as detected sources (not necessarily
GRB hosts) inside or on the border of the {\it Swift} XRT error circle
are concerned, they seem to be brighter (and apparently larger) for
bursts which have no detected OT. From our sample, the median AB
magnitude of sources inside the XRT error circle of such GRBs is 21.8,
whilst the corresponding value for the galaxies hosting GRBs
\emph{with} optical afterglows is 23.6 mag. This is admittedly small
number statistics, but it could be an indication that galaxies hosting
GRBs with very faint or even missing OTs belong to a different
population (i.e.\ more dusty galaxies with more homogeneously
distributed star-formation) than galaxies hosting GRBs with optical
afterglows. (We avoid referring to GRBs without a detected OT as
``dark'' bursts, as there are no generally accepted criteria for what
constitutes an optically dark GRB.)

Our Monte Carlo simulations indicate that the host galaxy sample
presented here (i.e.\ the 15 GRB fields corresponding to long, {\it
  Swift}-detected, OT-confirmed GRBs) is fainter than the pre-{\it
  Swift} sample in \citet{Fruchter06}. A possible explanation for this
is that they are more distant than the hosts in
\citeauthor{Fruchter06} This is in accordance with the observation
that {\it Swift} GRBs are at higher average redshift than pre-{\it
  Swift} bursts \citep[e.g.][]{Jakobsson06a}.  Consequently, GRBs
detected by the {\it Swift} satellite provide a better insight into
the early history of the Universe. 
The faintness of the GRB host galaxies calls for more extensive
observation efforts with large 8m class or space telescopes in order
to construct any representative sample of GRB host galaxies
\citep[e.g.][]{Jakobsson06b}.

\acknowledgments
We are grateful to the staff at ESO/La Silla for help with the control
system at the Danish 1.54m telescope and to Brian Lindgren Jensen at the
Dark Cosmology Centre for software support during the observing runs.



\end{document}